\documentclass{article}
\usepackage{times}
\usepackage{latexsym}

\usepackage{amsmath}
\usepackage{amsfonts}
\usepackage{amssymb}
\usepackage{amsthm}
\usepackage{url}
\usepackage[ruled,linesnumbered,vlined,scleft]{algorithm2e}
\usepackage{stmaryrd}

\newif\ifdraft
\drafttrue

\ifdraft
\usepackage{xcolor}
\newcommand{\comment}[1]{{\color{blue}\bf *** #1 ***}}
\else
\newcommand{\comment}[1]{}
\fi

\newtheorem{theor}{Theorem}[section]

\newtheorem{proposition}[theor]{Proposition}
\newtheorem{example}{Example}[section]
\newtheorem{remark}{Remark}

\newcommand{\nd}{\noindent}
\newcommand{\ie}{{\em i.e.}}
\newcommand{\eg}{{\em e.g.}}
\newcommand{\wrt}{{w.r.t.}}
\newcommand{\ii}[1]{\mbox{$(#1)$}}

\newcommand{\K}{\mathcal{K}}
\newcommand{\I}{\mathcal{I}}
\newcommand{\T}{\mathcal{T}}

\newcommand{\s}{\mathcal{S}}
\newcommand{\ro}{\mathcal{R}}
\newcommand{\D}{\mathcal{D}}
\newcommand{\A}{\mathcal{A}}
\newcommand{\R}{\mathcal{R}}
\newcommand{\C}{\mathcal{C}}
\newcommand{\E}{\mathcal{E}}
\newcommand{\at}{\mathcal{A}t}
\newcommand{\calc}{\mathtt{C}}
\newcommand{\x}{\mathtt{X}}
\newcommand{\calI}{\mathcal{I}}
\newcommand{\calP}{\mathtt{P}}
\newcommand{\abox}{\textsl{ABox}}
\newcommand{\tbox}{\textsl{TBox}}
\newcommand{\rbox}{\textsl{RBox}}
\newcommand{\dbox}{\textsl{DBox}}
\newcommand{\calO}{\mathcal{O}}

\newcommand{\tuple}[1]{\langle #1 \rangle}

\newcommand{\alc}{$\mathcal{ALC}$}

\newcommand{\shoin}{$\mathcal{SHOIN}$}
\newcommand{\sroiq}{$\mathcal{SROIQ}$}

\newcommand{\andc}{\sqcap}

\newcommand{\notc}{\neg}

\newcommand{\orc}{\sqcup}

\newcommand{\topc}{\top}

\newcommand{\impc}{\sqsubseteq}

\newcommand{\cond}{\stackrel{\sqsubset}{\scriptstyle\sim}}

\newcommand{\KB}{\mathcal{K}}
\newcommand{\self}{\mathtt{Self}}

\newcommand{\mat}[1]{\overline{#1}}

\newcommand{\compRank}{\ensuremath{\mathsf{ComputeRanking}}}

\newcommand{\dland}{\sqcap}
\newcommand{\dlor}{\sqcup}
\newcommand{\subs}{\sqsubseteq}

\newcommand{\dsubs}{\cond}
\newcommand{\entails}{\models}
\newcommand{\proves}{\vdash}
\newcommand{\assigned}{:=}
\newcommand{\rat}{\ensuremath{\mathsf{rat}}}

\title{A Rational Entailment for Expressive Description Logics via  Description Logic Programs}
\author{Giovanni  Casini$^{1,2}$ \  \  Umberto Straccia$^1$ \\  \\
{\small  \begin{tabular}{cc}
$^1$ISTI - CNR & $^2$CAIR - University of Cape Town \\
Italy & South Africa \\ \\ 
\multicolumn{2}{c}{Email: name.surname\mbox{@}isti.cnr.it} \\
\end{tabular} }
}

\begin{document}
\maketitle

\begin{abstract}
Lehmann and Magidor's \emph{rational closure} is acknowledged as a landmark in the field of non-monotonic logics and it has also been re-formulated in the context of \emph{Description Logics} (DLs).

We show here how to model a rational form of entailment for expressive DLs, such as \sroiq, providing a novel  reasoning procedure that compiles a non-monotone  DL knowledge base into a \emph{description logic program} (dl-program).

\end{abstract}

\section{Introduction}


One of the main non-monotonic formalism, namely~Lehmann and Magidor's \emph{rational closure}~\cite{LM92},  is acknowledged as a landmark for non-monotonic reasoning due to its logical properties. Rational closure, that falls under the more general class of the rational entailment relations \cite{LM92}, has been proposed in the context of \emph{Description Logics} (DLs)~\cite{BCGN03}, starting from basic DLs, such as \alc~\cite{Britz11,CS10,Casini11,CasiniStraccia13,BritzEtAl21,GiordanoEtAl15,GiordanoGliozzi21}, and re-formulated for low-complexity DLs, as $\mathcal{EL}_{\bot}$ \cite{CasiniEtAl2018,GiordanoDupre2016}, as for  expressive ones, up to \sroiq~\cite{Bonatti19}. 

Here we show  an implementation of a rational entailment relation for an expressive DL such as \sroiq~\cite{HorrocksEtAl2006}. The main contribution of this paper is that we re-formulate the decision procedure for rational closure by compiling a non-monotone  DL knowledge base into a \emph{description logic program} (dl-program)~\cite{Eiter08}. Dl-programs have been proposed to combine DLs  with \emph{Answer Set Programming}~\cite{Gelfond91}, an established approach to implement non-monotonic reasoning for rule-based languages.
In this way our approach can be easily be implemented on top of existing reasoners supporting dl-programs, such as DLV~\footnote{\url{http://www.dlvsystem.com}.}.

We proceed as follows. In section \ref{prel} we briefly present the logical systems we will refer to in the definition of our method, which is worked out in section \ref{rcdlp}. Eventually, in section~\ref{relw} we briefly recall related work and then we conclude.

\section{Preliminaries}\label{prel}

\nd For the sake of completeness and ease the reading, we introduce here a minimum of basic notions.

\subsection{Description logic programs}\label{dlps}

\paragraph{Normal logic programs.}
Assume a first-order vocabulary $\Phi=\tuple{\calP,\calc}$, with $\calc$ a set of constants $\{a,b,\dots\}$ and $\calP$ a set of predicates $\{p,q,\dots\}$, and let $\x$ be a set of variables $\{x,y,\dots\}$.
A \emph{term} $t$ is either a variable from $\x$ or a constant from  $\calc$, and an \emph{atom} is an expression $p(t_1,\ldots,t_n)$, where $p$ is a $n$-ary predicate in $\calP$ and each $t_i$ is a term. A \emph{literal} $l$ is an atom or its negation (via connective $\neg$), while a negation-as-failure literal (\emph{NAF-literal}) is of the form $not~l$, where $l$ is a literal. A \emph{rule} $r$ is an expression of the form ($m\geq k\geq 0$)

\vspace{-0.5cm}
\begin{equation}\label{rule}
a\leftarrow b_1,\ldots,b_k,not~b_{k+1},\ldots,not~b_m \ ,
\end{equation}
\nd where $a,b_1,\ldots,b_m$ are literals. Intuitively, a rule has to be read as `if we know that $b_1,\ldots,b_k$ are true, but we are not aware that $b_{k+1},\ldots,b_m$ are true, then we can conclude $a$'.
We indicate by $H(r)$ (head of $r$) the literal $a$, by $B^+(r)$ (positive body of $r$) the set $\{b_1,\dots,b_k\}$ and by $B^-(r)$ (negative body of $r$) the set $\{b_{k+1},\dots,b_m\}$. A \emph{normal program} $P$ is a finite set of rules, while a \emph{positive program} $P$ is a finite set of rules in which $B^-(r)=\emptyset$ for every rule $r$.

As usual, atoms, literals, rules and programs are considered \emph{ground} if they do not contain any variable. The \emph{Herbrand Universe} of a program $P$ ($HU_P$) is the set of all the constants that appear in $P$, while the \emph{Herbrand Base} of $P$ ($HB_P$) is the set of all the literals that can be constructed from the predicates in $P$ and the constants in $HU_P$. A ground instance of a rule $r$ is obtained substituting every variable occurring in $r$ with a constant symbol in $HU_P$, and, given a program $P$, $ground (P)$ is the set of all ground instances of rules in $P$.

From the semantics point of view, an interpretation $I$ of a program $P$ is a \emph{consistent} subset of $HB_P$, \ie~$I\subseteq HB_{P}$ and there is no atom $a$ such that both $a$ and $\neg a$ are in $I$. The truth value of a literal $l$ is true, false, or unknown in $I$ iff, respectively, $l\in I$, $\neg l \in I$, or $\{l,\neg l\}\cap I=\emptyset$, where $\neg \neg a$ is $a$. The satisfiability of a program $P$ is reduced to the satisfiability of its rules expressed in ground form: that is, $I$ is a \emph{model} of a program $P$ iff it is a model of $ground(P)$, \ie~ if $B^+(r)\subseteq I$ and $B^-(r)\cap I=\emptyset$, then  $H(r)\in I$ for every rule in $ground(P)$. 

In case of a positive program $P$, the \emph{answer set} of $P$ is the least model of $P$ with respect to set inclusion: the fact that $P$ is positive guarantees the uniqueness of its answer set~\cite{Eiter08}. If  $P$ is not positive, the notion of \emph{answer set} is defined via the so-called \emph{Gelfond-Lifschitz transformation} (see, \eg~\cite{Gelfond91}). Specifically, consider a program $P$ and an interpretation $I\subseteq HB_P$. The \emph{Gelfond-Lifschitz transformation} of $P$ relative to $I$ gives back a positive program $P^I$, and it is obtained from $ground (P)$ with the following procedure:

\vspace{-0.4cm}
\begin{itemize}
\item delete from $ground(P)$ every rule $r$ s.t. $B^-(r)\cap I\neq\emptyset$; and
\item from the remaining rules delete the negative part of the body.
\end{itemize}

\vspace{-0.4cm}
\nd In this way, we end up with a positive program $P^I$, and $I$ is an \emph{answer set} for $P$ iff $I$ is the answer set for the positive ground program $P^I$. We indicate with $ans(P)$ the set of the answer sets of a program $P$. Eventually, we  define a \emph{cautious} (resp., \emph{brave}) consequence relation $\models_{c}$ ($\models_{b}$) as follows: $P\models_{c} l$ ($P\models_{b} l$) iff the literal $l$ is true in any (some) answer set of $P$.

\begin{example}
Let $P$ a program composed of the following rules:

\vspace{-0.4cm}
\begin{eqnarray*}
feline(a) & \leftarrow & \\
feline(b)& \leftarrow & \\
big(b)& \leftarrow & \\
docile(x)& \leftarrow & feline(x),not~big(x) \ .
\end{eqnarray*}

\vspace{-0.2cm}
\nd Consider the interpretation $I=\{feline(a), feline(b), big(b), docile(a)\}$. Then $P^I$ is defined as follows:

\vspace{-0.5cm}
\begin{eqnarray*}
feline(a) & \leftarrow & \\
feline(b)& \leftarrow & \\
big(b)& \leftarrow & \\
docile(a)& \leftarrow & feline(a) \ .
\end{eqnarray*}

\vspace{-0.2cm}
\nd It is straightforwardly verified that $I$ is the least model of $P^{I}$ and, thus, $I$ is an (indeed, it's unique) answer  set of $P$.

\end{example}

\vspace{-0.2cm}
\paragraph{Description logics.}
We shall refer here to an expressive DLs, namely \sroiq~(for more details about it, we refer the reader to~\cite{HorrocksEtAl2006}). The \sroiq~signature is composed of a set of \emph{concept names} $\at=\{A, B,\ldots\}$, a set of \emph{role names} $\s=\{R,S,\dots\}$, and a set $\calO$ of \emph{individuals} $\{a,b,\dots\}$. The set of roles is $\ro=\s\cup\{R^-\mid R\in\s\}\cup\{U\}$, where $R^-$ is the inverse of a role $R$ ($R^{--}$ is $R$) and $U$ is the universal role. We can also compose the roles in $\ro$ into finite chains such as $R_1\circ\ldots\circ R_n$. The set $\C$ of \sroiq~concepts is defined inductively as: 


\begin{itemize}
    \item[(i)] $\at\subseteq \C$; 
    \item[(ii)] $\top,\bot\in\C$;
    \item[(iii)]  if $\{a_1,\ldots, a_n\}\subseteq \calO$, then $\{a_1,\ldots, a_n\}\in \C$;
    \item[(iv)]  if $C,D\in \C$, then $C\andc D, C\orc D, \notc C \in \C$; 
    \item[(v)]  if $C\in \C, R\in\ro$, then $\exists R.C, \forall R.C, \geq_n R.C, \leq_n R.C, \exists R.\self \in \C$.
\end{itemize}


\nd Condition (iii) indicates that the enumerated sets of individuals (\emph{nominals}) can be used also in the TBox as concepts. An interpretation is a pair $\tuple{\Delta^{\calI},\cdot^{\calI}}$, where $\Delta^{\calI}$ is a nonempty set, called \emph{domain}, 
and the \emph{interpretation function} $\cdot^{\calI}$ assigns to every individual a member of the domain $\Delta^{\calI}$, to every concept name a subset of $\Delta^{\calI}$, and to every role name a  subset of $\Delta^{\calI} \times \Delta^{\calI}$. The function $\cdot^{\calI}$ is extended to all the concepts and roles in the following way:
\begin{itemize}
\item $\{o_1,\ldots, o_n\}^{\calI}= \{o_1^{\calI},\ldots, o_n^{\calI}\}$;
\item $(C\andc D)^{\calI}= C^{\calI}\cap D^{\calI}$;
\item $(C\orc D)^{\calI}= C^{\calI}\cup D^{\calI}$;
\item $(\notc C)^{\calI}= \Delta^{\calI}/ C^{\calI}$; 
\item $(\exists R.C)^{\calI}= \{x\in\Delta^{\calI}\mid \exists y.(x,y)\in R^{\calI}\wedge y\in C^{\calI}\}$;
\item $(\forall R.C)^{\calI}= \{x\in\Delta^{\calI}\mid \forall y.(x,y)\in R^{\calI}\rightarrow y\in C^{\calI}\}$;
\item $(\geq_n R.C)^{\calI}= \{x\in\Delta^{\calI}\mid \#\{ y\mid (x,y)\in R^{\calI}\wedge y\in C^{\calI}\}\geq n\}$;
\item $(\leq_n R.C)^{\calI}= \{x\in\Delta^{\calI}\mid \#\{ y\mid (x,y)\in R^{\calI}\wedge y\in C^{\calI}\}\leq n\}$;
\item $(\exists R.\self)^{\calI}= \{x\in\Delta^{\calI}\mid (x,x)\in R^{\calI}\}$;
\item $(R^-)^{\calI} =\{(a,b)\mid (b,a)\in R\}$;
\item $(U)^{\calI} =\Delta^{\calI} \times \Delta^{\calI}$;
\item $(R_1\circ \ldots \circ R_n)^{\calI}=\{(a,b)\mid \exists x_1,\ldots, x_{n-1}.(a,x_1)\in R_1^{\calI},(x_1,x_2)\in R_2^{\calI},\ldots$, $(x_{n-1},b)\in R_n^{\calI}\}$.
 \end{itemize}


\nd where $\#S$ is the cardinality of set $S \subseteq \Delta^{\calI}$. A \emph{DL knowledge base} $L$ is a triple $\tuple{\A,\T,\R}$, where $\A$ is an \abox , containing information about the individuals,  $\T$ is a \tbox,  containing information about the relations between the concepts, and $\R$ is a \rbox, containing information about the roles.
The kind of axioms contained in the \abox, the \tbox, and the \rbox~are described in the table below, with their respective semantics ($n \geq 1$). 

%
%
%
\begin{center}
  {\footnotesize
  \begin{tabular}{@{} |c|c|c|c|  @{}}

\hline
 &   Axiom Name & Sintax & Semantics \\ 
\hline

 \textbf{ABox} &   Concept membership axiom & $C(a)$ & $a^{\I} \in C^{\I}$ \\ 
   & Role membership axiom & $R(a,b)$ & $(a^{\I},b^{\I}) \in R^{\I}$ \\ 
\hline

   \textbf{TBox} &   Concept inclusion axiom & $C\impc D$ & $C^{\I} \subseteq D^{\I}$ \\

\hline

   \textbf{RBox} &   
     Role inclusion axiom & $R_1\circ\ldots\circ R_n\impc S$ & $(R_1\circ\ldots\circ R_n)^{\I} \subseteq S^{\I}$ \\ 
   &  Transitivity & $Trans(R)$ & $R^{\I}$ is transitive \\ 
   &  Functionality & $Fun(R)$ & $R^{\I}$ is a function \\
   &  Reflexivity & $Ref(R)$ & $R^{\I}$ is reflexive \\
   &  Irreflexivity & $Irr(R)$ & $R^{\I}$ is irreflexive \\
   &  Simmetry & $Sym(R)$ & $R^{\I}$ is symmetric \\
   &  Asimmetry & $Asy(R)$ & $R^{\I}$ is asymmetric \\
   &  Disjointness & $Dis(R,S)$ & $R^{\I}$ and $S^{\I}$ are disjoint \\

\hline
  \end{tabular}
  }
  \end{center}
%

\nd A RBox has further to comply with an additional syntactical restriction: that is, a RBox has to be \emph{regular}, which essentially prevents a RBox from containing cyclic dependencies among roles that are known to lead to undecidability~\cite{Horrocks04b}. For ease of presentation we do not include the definition here and refer the reader to~\cite[Definition 2]{HorrocksEtAl2006} instead.
 We use $C = D$ as a shorthand of the concept inclusion axiom $\topc \impc (\notc C \orc D) \andc (\notc D \orc C)$. With $\models$ we denote the classical, monotonic, consequence/entailment relation, which is defined as usual.

Note also that every ABox axiom can be reformulated as an equivalent TBox axiom. In particular, $C(a)$ can be reformulated as $\{a\}\subs C$, while $R(a,b)$ is equivalent to $\{a\}\subs \exists R.\{b\}$. Consequently, in what  follows we will not consider ABoxes.

\paragraph{Description logic programs.}
A \emph{description logic program} (\emph{dl-program}) is composed of a pair $\KB=\tuple{L,P}$, where $L$ is a DL knowledge base and $P$ is a set of \emph{dl-rules}~\cite{Eiter08}, which we are going to specify next. The DL knowledge base $L$ is defined over a vocabulary composed of a set of concept names $\at$, a set of role names $\s$, and a set $\calO$ of individuals, while $P$ is defined over a vocabulary $\Phi=\tuple{\calP,\calc}$, with $\calc$ a set of constants and $\calP$ a set of predicates, and with $\x$ a set of variables. We assume that the predicative part of the two formalisms are independent, that is $\at\cup\s$ is disjoint from $\calP$, while the same domain of individuals is shared, that is $HU_P\subseteq \calc\subseteq \calO$. 

Dl-programs use the notions of  \emph{dl-query} and \emph{dl-atom} to be used in rule bodies to express queries to the DL knowledge base $L$. That is, a \emph{dl-query} $Q(\mathbf{t})$ can have various forms, but to what concerns us, it is sufficient to consider the following ones:

\begin{itemize}
\item a concept membership axiom $C(t)$ (so, $\mathbf{t}=t$);
\item a role membership axiom $R(t_1,t_2)$ (so, $\mathbf{t}=\tuple{t_1,t_2}$).
\end{itemize}


\nd On the other hand, a \emph{dl-atom} is an expression of form\footnote{The definition given here is again simpler than the original one, as we consider only the form strictly required for our proposal.}
\[
DL[S_1 \uplus p_1,\ldots,S_m \uplus p_m;Q](\mathbf{t})
\]
\nd with $m\geq 0$, where each $S_i$ is either a concept or a role ($S_i\in\C\cup\ro$), and each $p_i$ is a predicate symbol from $\calP$, unary if $S_i$ is a concept, binary otherwise, and $Q(\mathbf{t})$ is a dl-query.
%
%
The operator $\uplus$ is functional to the updating of the DL knowledge base $L$ with factual information obtained from the activation of the rules in the program. That is, each $S_i \uplus p_i$ indicates that the extension of $S_i$ is increased by the extension of $p_i$.

Now, a \emph{dl-rule} $r$ is of the form (\ref{rule}), where any literal $b_1,\ldots,b_m\in B(r)$ may be a dl-atom and 
a \emph{dl-program} is a pair $\KB=\tuple{L,P}$, where $L$ is a DL knowledge base and $P$ is a set of dl-rules.


From a semantics points of view, for an interpretations $I\subseteq HB_P$, 
we say that $I$ is a \emph{model} of a ground literal or dl-atom $l$ under $L$ ($I\models_L l$) iff
\begin{itemize}
\item if $l\in HB_P$, then $I\models_L l$ iff $l\in I$;
\item if $l$ is a ground dl-atom $DL[\lambda,Q](\mathbf{c})$, where $\lambda=S_1\uplus p_1,\ldots,S_m\uplus p_m$, then $I\models_L l$ iff $L(I;\lambda)\models  Q(\mathbf{c})$, where $L(I;\lambda)=L\cup\bigcup^m _{i=1} A_i(I)$, with, for $1\leq i\leq m$, $A_i(I)= \{S_i(\textbf{e})\mid p_i(\mathbf{e})\in I\}$,
%
%
\end{itemize}
 


As usual, an interpretation $I$ is a \emph{model} of a ground dl-rule $r$ iff $I\models_L l$ for all $l\in B^+(r)$ and $I\not\models_L l$ for all $l\in B^-(r)$ implies $I\models_L H(r)$. $I$ is a \emph{model} of a dl-program $\KB=\tuple{L,P}$ (written $I\models \KB$) iff $I\models_L r$ for all $r\in ground(P)$. We say that $\KB$ is \emph{satisfiable} if it has a model.


Let $KB=\tuple{L,P}$ be a dl-program. The \emph{strong dl-transform} of $P$ \wrt~$L$ and $I$ (denoted $sP^I _L$) is the set of all dl-rules obtained from $ground(P)$ by deleting
\begin{itemize}
\item every dl-rule $r$ s.t. $I\models_L l$ for some $l\in B^-(r)$; 
\item from the remaining dl-rule $r$ all literals in $B^-(r)$.
\end{itemize}

\nd Note that \ii{i} $\tuple{L,sP^I _L}$ has only monotonic dl-atoms and no NAF-literals anymore; and \ii{ii} a \emph{positive} dl-program, if satisfiable, has a least model~\cite{Eiter08}. Now, a \emph{strong answer set} of $\KB=\tuple{L,P}$ is an interpretation $I\subseteq HB_P$ s.t. $I$ is the least model of $\tuple{L,sP^I _L}$. We denote with $ans_s (\KB)$ the set of the strong answer sets of $\KB$. $l$ is a \emph{cautious} (\emph{brave}) consequence of $\KB$, indicated as $\KB\models_{s,c} l$ ($\KB\models_{s,b} l$) iff $l$ is true in every (some) strong answer of $\KB$.

Note that given a dl-program $\KB=\tuple{L,P}$ and an answer set  $I$ of $\KB$, $I$ is a minimal model of $\KB$~\cite{Eiter08}.

%

\begin{example}\label{dl-prog}
Consider a dl-program $\KB=\tuple{L,P}$. Let $L=\tuple{\T}$, with 
{\footnotesize
\[
\T=\{\{a\}\subs Cat, \{b\}\subs Feline, \{b\}\subs Big\} \ , 
\]
}
 and consider a dl-program $P$ composed of the following rules:
{\footnotesize
\begin{eqnarray*}
feline(x) & \leftarrow  & DL[Cat](x) \\
docile(x)& \leftarrow  & DL[Feline\uplus feline;Feline](x),  not~DL[Big](x) \ .
\end{eqnarray*}
}
\nd It can easily be shown that $\KB$ has an unique answer set 
{\footnotesize
\[
I=\{feline(a), docile(a)\} \ .
\]
}
\nd In fact, $I$ is the  least model of the following set $sP^I _L$ of dl-rules:
{\footnotesize
\begin{eqnarray*}
feline(b) & \leftarrow  & DL[Cat](b) \\
feline(a) & \leftarrow  & DL[Cat](a) \\
docile(a) & \leftarrow  & DL[Feline\uplus feline;Feline](a) \ .
\end{eqnarray*}
}
\end{example}

\subsection{Rational Closure for \alc}\label{rc}

For convenience, we recap here some salient notions related to \emph{rational closure} (RC) for DLs, specifically for the DL~\alc~ (see, \eg~\cite{Casini10}).

\begin{remark}
We remind that \alc~concepts are inductively defined as
(i) $\at\subseteq \C$; 
(ii) $\top,\bot\in\C$;
(iii)  if $C,D\in \C$, then $C\andc D, C\orc D, \notc C \in \C$; 
(iv)]  if $C\in \C, R\in\ro$, then $\exists R.C, \forall R.C \in \C$. 
\end{remark}


\nd Now, a \emph{defeasible concept inclusion} axiom is of the form $C \cond D$, where, without loss of generality, $C$ and $D$ are assumed to be  atomic concepts or their negation. 
%
%
The expression $C\cond D$ has to be read as `if an individual falls under the concept $C$, typically it falls also under the concept $D$'.   A \emph{defeasible} DL knowledge base is a pair $L=\tuple{\T,\D}$, where $\T$ (the \tbox) is a finite set of concept inclusion axioms of the form $C\impc D$, where $C,D$ are \alc~concepts, and $\D$ (the $\dbox$) is a finite set of defeasible concept inclusion of the form $C\cond D$.

We next briefly describe  the decision procedure for RC for \alc, referring in particular to the one presented in \cite{BritzEtAl21}, that in turn has been obtained by refining the one presented in \cite{Casini10}. 
%
%
Consider $L=\tuple{\T,\D}$. The first step of the procedure is to assign a rank to each defeasible axiom in~$\D$. The rank of the defeasible axioms indicates, in case of conflictual information, which axiom is associated to more specific premises, and has the priority over the axioms associated to more general premises. Central to this step is the exceptionality procedure $\mathtt{Exceptional}(\cdot)$ (see below). The procedure makes use of the notion of \emph{materialisation}, to reduce concept exceptionality checking to entailment checking, were the  \emph{materialisation} of~$\D$ is defined as $
\mat{\D}:=\{\lnot C\dlor D \mid C\cond D\in\D\}$.

\begin{procedure}[ht]
\KwIn{A DL knowledge base $L=\tuple{\T,\D}$}
\KwOut{$\E\subseteq\D$}
$\E\assigned\emptyset$\;
\ForEach{$C\dsubs D\in\D$}{	
   	\If{$\T\entails\bigsqcap\mat{\D}\subs\lnot C$}{$\E\assigned\E\cup\{C\dsubs D\}$}
 }
\textbf{return} $\E$
\caption{Exceptional($L$)}\label{Func:Exceptional}
\end{procedure}

\nd The ranking of the defeasible axioms is done via the $\mathtt{ComputeRanking}(\cdot)$ procedure.
\begin{procedure}[ht]
\caption{ComputeRanking($L$)\label{Func:Ranking}}
\KwIn{A DL knowledge base $L=\tuple{\T,\D}$}
\KwOut{$L^*=\tuple{\T^*,\D^*}$ and an exceptionality ranking $\E$}
$\T^*\assigned\T$\;
$\D^*\assigned\D$\;
\Repeat{$\D^*_{\infty}=\emptyset$}
	{
	$i\assigned 0$\;
	$\E_{0}\assigned\D^*$\;
	$\E_{1}\assigned \text{Exceptional}(\tuple{\T^*,\E_{0}}$)\;
	\While{$\E_{i+1}\neq\E_{i}$}{
		$i\assigned i + 1$\;
		$\E_{i+1}\assigned \text{Exceptional}(\tuple{\T^*,\E_{i}}$)\;
	}
	$\D^*_{\infty}\assigned\E_{i}$\;
	$\T^*\assigned\T^*\cup\{C\subs D\mid C\dsubs D\in\D^{*}_{\infty}\}$\;
	$\D^*\assigned\D^*\setminus\D^*_{\infty}$\;
	}
	$\E\assigned (\E_0,\ldots,\E_{i-1})$\;
\textbf{return} ($L^*=\tuple{\T^*,\D^*}$, $\E$)\;
\end{procedure}
In short, the $\mathtt{ComputeRanking}(\cdot)$ takes as input $L=\tuple{\T,\D}$ and gives back a new semantically equivalent knowledge base $L=\tuple{\T^*,\D^*}$ (with $\T\subseteq\T^*$ and $\D^*\subseteq\D$), where possibly some defeasible information in $\D$ has been identified as strict and added to $\T$. Also, a sequence of $\subseteq$-ordered subsets of $\D$ $(\E_0,\ldots,\E_{i-1})$, with increasing level of specificity. That is, in case of potential conflicts, the axioms in a set $\E_j$, $j\geq 0$, have the priority over the axioms in any $\E_i$, $0\leq i< j$. 
Now, by considering the ranking $\E_0,\ldots,\E_{i-1}$, we can define a ranking function $r$ that associates to every defeasible concept inclusion in $\D$ a number, representing its level of exceptionality: that is,

\[
 r(C\cond D)=    \left\{\begin{array}{lll}
     j &\mathrm{if}\ C\cond D\in \mathcal{E}_j\ \mathrm{and}\ C\cond D\notin \mathcal{E}_{j+1}\\
     \infty&\mathrm{if}\ C\cond D\in \mathcal{E}_j\ \mathrm{for\ every}\ j \ .
     \end{array}\right.
\]

\nd Similarly, we may associate a rank to a concept $C$ in the following way: consider the result $(L^{*}=\tuple{\T^{*},\D^{*}},\E=(\E_0,\ldots,\E_n))$ of 
$\mathtt{ComputeRanking}(\cdot)$. Then 

\[
 r(C)=    \left\{\begin{array}{lll}
     j &\mathrm{if} \  \T^{*}\entails\bigsqcap\mat{\E}_{j}\subs\notc C \ \mathrm{and}\ \T^{*}\not\entails\bigsqcap\mat{\E}_{j+1} \subs\notc C \ \\
     \infty&\mathrm{if }\ \T^{*}\entails\bigsqcap\mat{\E}_{j} \subs\notc C \ \mathrm{for\ every}\ j \ .
     \end{array}\right.
\]

\nd Note that $ r(C\cond D) = r(C)$. Now, we will say that $C\cond D$ is \emph{entailed} by the rational closure of a DL knowledge base $L$ (denoted $L \proves_{\rat}C\dsubs D$) iff $r(C)<r(C\andc \neg D)$~\cite[Theorem 5.17]{LM92}).
Finally, the procedure $\mathtt{RationalClosure}(\cdot)$ determines whether $L\proves_{\rat}C\dsubs D$.
We recall that the defined entailment relation is indeed as so-called \emph{rational consequence relation}~\cite{LM92}, \ie~satisfies the following properties:\\

\begin{footnotesize}
 \begin{tabular}{lll}
     $\mathrm{(REF)}$ & \ \ $L \proves_{\rat} C\cond C$ & Reflexivity \\\\
     
     $\mathrm{(LLE)}$&$\frac{\begin{array}{lcr}
         L \proves_{\rat} C\cond F && L \models C = D\\
         \end{array}}{\begin{array}{lcr}
         &L \proves_{\rat} D\cond F&\\
         \end{array}}$&Left Logical Equival.\\\\

        $\mathrm{(RW)}$&$\frac{\begin{array}{lcr}
         L \proves_{\rat} C\cond D && L \models D\subs F\\
         \end{array}}{\begin{array}{lcr}
         &L \proves_{\rat} C\cond F&\\
         \end{array}}$&Right Weakening\\\\
         
     $\mathrm{(CT)}$ & $\frac{\begin{array}{lcr}
         L \proves_{\rat} C\cond D && L \models C \andc  D \cond F\\
         \end{array}}{\begin{array}{lcr}
         &L \proves_{\rat} C\cond F&\\
         \end{array}}$& Cut (Cumulative Trans.)\\\\


     $\mathrm{(OR)}$& $\frac{\begin{array}{lcr}
         L \proves_{\rat} C\cond F && L \proves_{\rat} D \cond F\\
         \end{array}}{\begin{array}{lcr}
         &L \proves_{\rat} C\orc D \cond F&\\
         \end{array}}$&Left Disjunction\\\\

     $\mathrm{(RM)}$& $\frac{\begin{array}{lcr} L \proves_{\rat} C\cond F && L \not \proves_{\rat} C\cond \neg D\end{array}}
         {\begin{array}{c} L \proves_{\rat} C\andc D \cond F \end{array}}$ & Rational Monotony\\\\

\end{tabular}
\end{footnotesize}

%
%

\begin{procedure}[ht]
\caption{RationalClosure($L$, $\alpha$)\label{Func:RationalClosure}}
\KwIn{$L=\tuple{\T,\D}$ and a query $\alpha=C\dsubs D$.}
\KwOut{$\mathtt{true}$ if $L \proves_{\rat}C\dsubs D$, $\mathtt{false}$ otherwise}
$(L^{*}=\tuple{\T^{*},\D^{*}},\E=(\E_0,\ldots,\E_n))\assigned \compRank(L)$\;
$i\assigned 0$\;
\While {$\T^{*}\entails\bigsqcap\mat{\E_{i}}\dland C\subs\bot$ and $i\leq n$}
{$i\assigned i + 1$\;
}
\If{$i\leq n$}
	{\textbf{return} $\T^{*}\entails\bigsqcap\mat{\E_{i}}\dland C\subs D$\;}
\Else{\textbf{return} $\T^{*}\entails C\subs D$\;}
\end{procedure}

\nd We refer the reader to \cite{BritzEtAl21} for further explanations and details and limit our presentation to a concluding example.

\vspace{-0.6cm}
\begin{example} \label{exA}
Assume a DL knowledge base $\tuple{\T,\D}$ with 
\begin{eqnarray*}
\T & = \{& 
Cat \subs  Feline, Tiger\subs Feline, Tiger\subs Big,  BigFeline = Feline\andc Big \ \ \}\\
\D & =\{& Feline\cond Agile, Feline\cond Docile, BigFeline \cond \neg Docile \ \ \} \ . 
\end{eqnarray*}
\nd By applying the ranking procedure, we end up with 
\[
\begin{array}{l}
r(Cat)  =  r(Feline) = 0 \\
r(Feline\cond Agile)  =  r(Feline\cond Docile)=0 \\ \\
r(Tiger)  =  r(Feline\andc Big)=1\\
r(BigFeline \cond \neg Docile)  =  1 \ . 
\end{array}
\]

\nd So, for instance, we can conclude that 
\[
\begin{array}{l}
\KB  \proves_{\rat}  Cat\cond Docile \ , 
\KB  \proves_{\rat} Cat\cond Agile \ ,
\KB  \proves_{\rat}  Cat\cond \neg Big \\
\KB  \proves_{\rat}  Tiger\cond \neg Docile \ , 
\KB  \proves_{\rat}  Cat\cond \neg Tiger \ .
\end{array}
\]
\end{example}



\section{Rational entailment for DLs via dl-programs}\label{rcdlp}

In this section we show that, starting from a non-monotone DL (\sroiq) knowledge base 
$L = \tuple{\T,\R,\D}$, we can compile $L$ into a dl-program $\K = \tuple{\tuple{\T^*,\R}, P}$ such that the conditions for rational consequence relations are preserved.
%
%
%
So, consider a defeasible \sroiq~ knowledge base $L = \tuple{\T,\D}$. Our approach consists of two steps: a ranking step and a compilation step.

\emph{Ranking step.}
To $L$ we apply  the procedure $\mathtt{ComputeRanking}(L)$ described in Section~\ref{rc}\footnote{Of course,   $\mathtt{Exceptional}(\cdot)$ and, thus, $\mathtt{ComputeRanking}(\cdot)$, can be applied to  a DL \sroiq~knowledge base $L$ as the classical entailment relation for \sroiq~is decidable.} and, thus, we end up with a new defeasible DL knowledge base $L^*=\tuple{\T^*, \R, \D^*}$ that  correctly separates the strict and the defeasible information contained in the original pair $L=\tuple{\T,\R,\D}$, and a ranking value $r(C\cond D)$ for every defeasible axiom $C\cond D \in \D^*$.

Note that in order to adapt the procedures $\mathtt{Exceptional}(\cdot)$ and $\mathtt{ComputeRanking}(\cdot)$ to \sroiq~it is sufficient to consider also $\R$ into the ranking procedure: the inputs of both the procedures is a DL knowledge base $L = \tuple{\T,\R,\D}$ instead of $L = \tuple{\T,\D}$, and line 3 in Procedure $\mathtt{Exceptional}(\cdot)$ is modified from $\T\entails\bigsqcap\mat{\D}\subs\lnot C$ to $\T\cup\R\entails\bigsqcap\mat{\D}\subs\lnot C$. The set $\R$ comes out untouched from the ranking procedure, since $\D$ is the only ranked set, and the only possible new strict information is of the form $C\subs D$, with $C$ and $D$ concepts, hence it can affect only the content of $\T$. Hence, starting from a knowledge base $\tuple{\T,\R,\D}$  we end up with a ranked knowledge base $\tuple{\T^*,\R,\D^*}$.

\vspace{-0.3cm}
\paragraph{Dl-program compilation step.}
Given $L^*=\tuple{\T^*,\R, \D^*}$ from the ranking step, we now compile the defeasible information in $\D^*$ into a a set of dl-rules $P$, which together with $\T^*,\R$ defines then the final dl-program $\K = \tuple{\tuple{\T^*,\R}, P}$. 

\emph{To alleviate the reading, let $L= \tuple{\T,\R,\D}:=\tuple{\T^*,\R,\D^*}$; that is, we assume that $\tuple{\T,\R,\D}$ has already been ranked via the previous ranking step.} Now, define a signature $\Phi=\tuple{\calP,\calc}$ with $\calc=\calO$, while $\calP$ is composed of the predicates ($c,d,e,\dots$) representing at the level of programs the concept names in $\T\cup\D$, \ie~for each concept $C$ in $\at$  we have an unary predicate $c$ representing it in the rules. We will use the same name with or without the uppercase initial letter to indicate if it is a concept in DL (\eg, $Male$) or a predicate in $P$~(\eg, $male$), respectively. Let us also recall that  for each $C\cond D \in \D$, $C$ and $D$ are either atomic concepts or their negation. Given the ranking of the defeasible axioms in  $\D$, let 
\[
\D_k =\{C\cond D\mid C\cond D\in\D\ \text{and}\ r(C\cond D)=k\}
\]
\nd be the subset of $\D$ composed of the axioms with rank value $k$. Now, define the set 
\[
\mathfrak{A}_{\D_k}=\{C\mid C\cond D\in \D_k\}
\]
\nd as the set of all the antecedents of the defeasible axioms of rank $k$. Moreover, we consider also the set of the consequents of the defeasible axioms 
\[
\mathfrak{C}_\D=\{D\mid C\cond D\in\D\} \ .
\]

\nd Now, for every axiom $C\cond D\in \D$ of rank $k$, we create a pair of rules of the form\footnote{We assume to simplify double negation: that is, for a concept name $F$, $\neg\neg F$ is $F$, and similarly, for a logic program predicate $f$,  $\neg\neg f$ is $f$. See also Example~\ref{exB} later on.}
%
%
%

\begin{eqnarray}
d(x) & \leftarrow & DL[\lambda;C](x), \nonumber \\
 && not~DL[\lambda;\bigsqcup \{C'\mid  C'\in\mathfrak{A}_{\D_m},\  \mathrm{with}\ m>k \}](x),  \nonumber \\
 && not~\neg d(x) \label{1} \\ \nonumber \\
 \neg d(x) & \leftarrow & DL[\lambda;\neg D](x) \ . \label{2}
\end{eqnarray}

\nd Additionally, for all $C\in\mathfrak{A}_{\D_m}$ with $m>1$, we also consider a rule
\begin{equation}\label{3}
\neg c(x)\leftarrow not\ DL[\lambda; C](x) \ . 
\end{equation}

\nd In all rules above, $\lambda=\{E\uplus\ e, \neg E\uplus\ \neg e\mid E\in \mathfrak{C}_\D\}$. 
Note that the size of the grounding of the compiled dl-program is polynomially bounded by the size of the defeasible DL knowledge base.

The intuitive meaning of the rule (\ref{1}) is the following: assume we have an individual $a$ that is an instance of concept $C$, which is the antecedent of the defeasible axiom $C\cond D$ of rank $k$; if $a$ is not an instance of any other $\D$-antecedent that is more exceptional than $C$, \ie~ $not~DL[\lambda;\bigsqcup \{C'\mid C' \in\mathfrak{A}_{\D_m},\  \mathrm{with}\ m>k \}](x)$ holds, and $d(a)$ is consistent with our knowledge base, then we can conclude $d(a)$. 

On the other hand, the purpose of rule (\ref{2}) is to update $P$, in case we  derive in $L$ that the conclusion of a defeasible axiom is negated and, thus, the defeasible axiom cannot be applied. $\lambda$ is necessary to update the DL-base $L$ with the conclusions drawn at the program level.

Finally, rules of form (\ref{3}) impose that the individuals we are dealing with are as typical as possible. That is, if we are not aware that an exceptional premise apply to them (any concept in $\mathfrak{A}_{\D_m}$, with $m>1$), then we assume that it doesn't apply (\eg, if we note that $a$ is a bird, but we are not aware that it is a penguin, then we presume that it is not a penguin). In the following, we illustrate our technique via an example.

\begin{example}\label{exB}

\nd Assume we have a DL vocabulary with  $\at = \{ B,P,F,I,Fi,W, Preyins$, $Preyfish \}$, $\s = \{Prey\}$, and 
$\calO = \{a,b\}$,
were the symbols stand for; $B \mapsto$ \emph{bird}, $P\mapsto$  \emph{penguin}, $F\mapsto$ \emph{flies}, $I\mapsto$ \emph{insect}, $Fi\mapsto$  \emph{fish}, $W\mapsto$  \emph{has wings}, $Preyins\mapsto$  \emph{eats insects}, $Preyfish\mapsto$ \emph{eats fishes}, while $Prey$ is the relation \emph{preys on}. 

The DL base $L=\tuple{\T,\D}$ is composed of 
\begin{eqnarray*}
\T & =\{ & \{a\}\impc B, \{b\}\impc P, P\impc B,I\impc\neg Fi, \\ && Preyins=\forall Prey.I\andc \exists Prey.\top, \\ && Preyfish=\forall Prey.Fi\andc \exists Prey.\top \ \} \\\\
\D & =& \{B\cond F, P\cond \neg F, B\cond Preyins, P\cond Preyfish, B\cond W \ \} \ .
\end{eqnarray*}

\nd Now, it can be verified that the ranking step returns the following ranking of axioms $\D$: 
\begin{eqnarray*}
\D_0 & =\{ & B\cond F, B\cond Preyins, B\cond W \ \} \\
\D_1 & =\{ & P\cond \neg F, P\cond Preyfish \ \} \ .
\end{eqnarray*}

\nd Therefore,  $\mathfrak{A}_{\D_0}  =\{  B \ \}$,  
$\mathfrak{A}_{\D_1}  =\{  P \ \}$,  
$\mathfrak{C}_\D  =\{  F, \neg F, Preyins, Preyfish \ \}$.
The compilation step proceeds now as follows.  We define a vocabulary $\Phi=\tuple{\calP,\calc}$ with $\calc=\{a,b\}$, while $\calP$ is composed of predicates that represent at the program level the DL atomic concepts and roles: that is, $\calP=\{b,p,f,i,fi,w, preyins,preyfish,prey\}$. The program $P$, resulting from the compilation step, is composed of the following rules:
{\footnotesize
\begin{eqnarray*}
f(x) & \leftarrow & DL[\lambda;B](x), not\ DL[\lambda;P](x), not\ \neg f(x) \\
 \neg f(x)& \leftarrow & DL[\lambda;\neg F](x) \\\\
 preyins(x)& \leftarrow & DL[\lambda;B](x), not\ DL[\lambda;P](x), not\ \neg preyins(x) \\
 \neg preyins(x)& \leftarrow & DL[\lambda;\neg Preyins](x) \\\\
 w(x)& \leftarrow & DL[\lambda;B](x), not\ DL[\lambda;P](x), not\ \neg w(x) \\
 \neg w(x)& \leftarrow & DL[\lambda;\neg W](x) \\\\
 \neg f(x)& \leftarrow & DL[\lambda;P](x), not\  f(x) \\
  f(x)& \leftarrow & DL[\lambda; F](x) \\\\
 preyfish(x)& \leftarrow & DL[\lambda;P](x), not\ \neg preyfish(x) \\
 \neg preyfish(x)& \leftarrow & DL[\lambda;\neg Preyfish](x) \\\\
 \neg p(x)& \leftarrow & not\ DL[\lambda; P](x) \ ,
\end{eqnarray*}
}
\nd with 
\begin{eqnarray*}
\lambda & =\{ & F\uplus f, \neg F\uplus \neg f, W\uplus w,\neg W\uplus \neg w, Preyins\uplus prey ins, \\
&& \neg Preyins\uplus \neg preyins, Preyfish\uplus preyfish, 
\neg Preyfish\uplus \neg preyfish \ \} \ .
\end{eqnarray*}


\nd Now, note that the only answer set to the program $P$ is the interpretation 
\[
I=\{f(a), preyins(a), w(a), \neg p(a), \neg f(b),  preyfish(b)\} \ .
\] 
\nd In fact, $I$ is the least model of the grounded positive program $P^I$
\begin{eqnarray*}
f(a) & \leftarrow & DL[\lambda;B](a) \\
preyins(a)& \leftarrow & DL[\lambda;B](a) \\
w(a)& \leftarrow & DL[\lambda;B](a)\\
\neg f(b)& \leftarrow & DL[\lambda;P](b)\\
preyfish(b)& \leftarrow & DL[\lambda;P](b) \\
\neg p(a)& \leftarrow &  \ .
\end{eqnarray*}


\nd So, we obtain the intuitive conclusions that, if we are aware about  an individual  that it is just a bird, we can conclude that, presumably, it flies, eats insects and has wings. On the other hand, if we are informed that it is a penguin, we can conclude that it doesn't fly and eats fishes. 
\end{example}


 
\nd As well known and already noted in~\cite{CS10}, having \emph{nominal concepts} may end up in having multiple extensions, \ie, in our context, we may have  multiple strong answer sets as shown with following simple example.

\begin{example}
Consider a knowledge base $L=\tuple{\T,\D}$, with
\begin{eqnarray*}
\T & = \{ & \{a\}\impc\exists R.\{b\}, C=D\andc \forall R.\neg D \ \} \\
\D & = \{ & \top\cond C \ \} \ .
\end{eqnarray*}

\nd By applying our method we obtain  the following program $P$
\begin{eqnarray*}
c(x)& \leftarrow & DL[\lambda;\top](x), not\ \neg c(x) \\
\neg c(x)& \leftarrow & DL[\lambda;\neg C](x) \ .
\end{eqnarray*}

\nd Now, it can be verified that from the dl-program $\KB=\tuple{L,P}$ we obtain now two strong answer sets: namely,
\begin{eqnarray*}
I & =\{ & c(a),\neg c(b) \ \} \\ 
I' & =\{ & c(b),\neg c(a) \ \}  \ . 
\end{eqnarray*}

\end{example}

\nd Nevertheless, the main result of this paper is that each strong answer set defines a rational consequence relation. 
In fact, we consider the content of the DL base updated with the content of an answer set $I$ by means of the operator $\uplus$. That is, we define a consequence relation $\models_{P^I}$ where, $\KB= \tuple{L,P}\models_{P^I}C(a)$ iff the DL base $L$ augmented, using $\uplus$, with the content of a strong answer set $I$ of $\KB$, entails $C(a)$. Specifically, we can show that

\begin{proposition}
Given $\KB=\tuple{L,P}$, were $L$ contains a \sroiq~\tbox~and a \sroiq~\rbox,  $P$ is the result of compiling $L$ into dl-rules, and a strong answer set $I$ of $\KB$. Then the consequence relation $\models_{P^I}$ satisfies the following properties:\footnote{For ease of comprehension, we write concept assertions as $D(b)$  in place of the equivalent inclusion axiom $\{b\} \subs D$ in expressions like $L\cup\{D(b)\}$.}

\begin{center}
\begin{footnotesize}
\begin{tabular}{ll}

    $\mathrm{REF_{DL}}$ & \ \ $\tuple{L,P}\models_{P^I} C(a) \ \mathrm{for\ every}\ C(a) \in L$ \\\\

    $\mathrm{LLE_{DL}}$&$\frac{\begin{array}{lcr}
        \tuple{L\cup \{D(b)\},P}\models_{P^I} C(a) && L \models D= E\\
        \end{array}}{\begin{array}{lcr}
        &\tuple{L\cup \{E(b)\},P}\models_{P^I} C(a)&\\
        \end{array}}$\\\\

    $\mathrm{RW_{DL}}$&$\frac{\begin{array}{lcr}
        \tuple{L,P}\models_{P^I} C(a)&& L \models C\sqsubseteq D\\
        \end{array}}{\begin{array}{lcr}
        &\tuple{L,P}\models_{P^I} D(a)&\\
        \end{array}}$\\\\

    $\mathrm{CT_{DL}}$ & $\frac{\begin{array}{lcr}
        \tuple{L\cup \{D(b)\},P}\models_{P^I} C(a) &&\tuple{L,P}\models_{P^I} D(b)\\
        \end{array}}{\begin{array}{lcr}
        &\tuple{L,P}\models_{P^I} C(a) &\\
        \end{array}}$\\\\


    $\mathrm{OR_{DL}}$& $\frac{\begin{array}{lcr}
        \tuple{L\cup \{D(b)\},P}\models_{P^I} C(a) &&\tuple{L\cup \{E(b)\},P}\models_{P^I} C(a)\\
        \end{array}}{\begin{array}{lcr}
        &\tuple{L\cup \{(D\sqcup E)(b)\},P}\models_{P^I} C(a) &\\
        \end{array}}$\\\\

    $\mathrm{RM_{DL}}$& $\frac{\begin{array}{lcr}\tuple{L,P}\models_{P^I} C(a) &&\tuple{L,P}\not\models_{P^I}
    \neg D(b)\end{array}}
        {\begin{array}{c}\tuple{L\cup \{D(b)\},P}\models_{P^I} C(a)\end{array}}$
    \end{tabular}

\end{footnotesize}
\end{center}

\end{proposition}

Due to space limits, here we omit the proof.


 \begin{proof}
 \nd (Sketch) The proofs for $\mathrm{REF_{DL}}$,  $\mathrm{LLE_{DL}}$ are $\mathrm{RW_{DL}}$ are straightforward, considering the set-theoretic semantics of DLs.

 In what follows, given the strong answer set $I$, the expression $I^{DL}$ indicates the obvious translation of the answer set into the DL base $L$, so that $\tuple{L,P}\models_{P^I} C(a)$ iff $L\cup I^{DL}\models C(a)$.

 For $\mathrm{CT_{DL}}$, if $I$ is an answer set for both $\tuple{L,P}$ and $\tuple{L\cup\{D(b)\},P}$, then we have $L\cup\{D(b)\}\cup I^{DL}\models C(a)$ and $L\cup I^{DL}\models D(b)$, and, since every classical DL consequence relation $\models$ satisfies $CT$, we have $L\cup I^{DL}\models C(a)$, \ie~$\tuple{L,P}\models_{P^I} C(a)$. 

 For $\mathrm{OR_{DL}}$, if $I$ is an answer set for both $\tuple{L\cup\{D(b)\},P}$ and $\tuple{L\cup\{E(b)\},P}$, it must be an answer set also for $\tuple{L\cup\{(D\orc E)(b)\},P}$: since $\models$ is monotonic, it is not possible to derive from $L\cup\{(D\orc E)(b)\}$ some element of the set $B^-(r)$ of some $r$ in $P$ that could not be derived from $L\cup\{D(b)\}$ or $L\cup\{E(b)\}$; hence a rule can be eliminated from $P$ only if also $L\cup\{D(b)\}$ or $L\cup\{E(b)\}$ would eliminate it. Given the validity of $OR$ for $\models$, we have that $L\cup\{D(b)\}\cup I^{DL}\models C(a)$ and $L\cup\{E(b)\}\cup I^{DL}\models C(a)$ imply $L\cup\{(D\orc E)(b)\}\cup I^{DL}\models C(a)$, \ie~$\tuple{L\cup \{D(b)\},P}\models_{P^I} C(a)$.

 For $\mathrm{RM_{DL}}$, assume $\tuple{L,P}\models_{P^I} C(a)$ and $\tuple{L,P}\not\models_{P^I}\neg D(b)$. It is sufficient to show that the answer set $I$ must be an answer set also for $\tuple{L\cup \{D(b)\},P}$. Assume the opposite, \ie~$I$ is not an answer set for $\tuple{L\cup \{D(b)\},P}$. Then, there must be in $P$ a rule $r$ associated to a defeasible axiom with rank equal to $k$ s.t. $not\ \alpha\in B^-(r)$, where $\alpha$ is some literal s.t. $L\not\models \alpha^{DL}$ and  $L\cup\{D(b)\}\models \alpha^{DL}$ ($\alpha^{DL}$ is the translation of $\alpha$ into the DL-language). In such a case, $r$ must have been a ground rule of form
 \[
 \begin{split}
 e(c)\leftarrow & DL[\lambda;C](c), not~DL[\lambda;\bigsqcup \{C'\mid\\ 
 & C' \in\mathfrak{A}_{\D_m},\  \mathrm{with}\ m>k \}](c), not \neg e(c) \ .
 \end{split}
 \]

 \nd $\alpha$ cannot be $\neg e(c)$, since from the activation of the rule we would have  $\tuple{L,P}\models_{P^I} e(c)^{DL}$, and consequently $\tuple{L,P}\models_{P^I} \neg D(b)$, which contradicts the hypothesis. As a consequence, 
 $\alpha$ must be the dl-atom of the form
 $not~DL[\lambda;\bigsqcup\{C' \}](c)$.
  But then again,  the activation of the rule for the individual $c$ under $\tuple{L,P}$ implies that the individual $c$ is ranked at the value $k$.\footnote{The rank of an individual $a$ is the rank of $\{a\}$, \ie~$r(\{a\})$.}
  %
  %
  Having every $C'$ a higher ranking value than $k$, and so also $\bigsqcup\{C'\}$, we can conclude $\tuple{L,P}\models_{P^I}\neg \bigsqcup\{C'\}(c)$, from which, again, we have $\tuple{L,P}\models_{P^I}\neg D(b)$, contrary to hypothesis. This concludes the proof.
 
 \end{proof}

\section{Related work} \label{relw}

Several non-monotonic DLs exist, but somewhat related to our proposal are~\cite{Britz11,Straccia93,CS10,Casini11,CasiniStraccia13,BritzEtAl21,GiordanoEtAl15,GiordanoGliozzi21,CasiniEtAl2018,GiordanoDupre2016,Bonatti19,PenselTurhan18,BritzV17,BritzV17a,GiordanoDupre2016,GiordanoDupre16a}, as they address the application of the preferential semantics \cite{LM92}. 
As far as we know, \cite{BritzV17,BritzV17a,Bonatti19} are the only works that consider also a DL as expressive as \sroiq. \cite{BritzV17,BritzV17a} propose a language, associated to a preferential semantics, that is more expressive than the one presented here, allowing the representation of many forms of defeasibily. However, at the moment such a logic is still missing a mature entailment relation. Bonatti \cite{Bonatti19} defines a semantic construction that extends rational closure to \sroiq: the previous proposals \cite{GiordanoEtAl15,CasiniEtAl2018,BritzEtAl21} rely on the \emph{disjoint model union property}, that does not hold for a DL as expressive as \sroiq, while Bonatti proposes an alternative construction based on \emph{stable rankings}, that is applicable for every DLs. We are not aware of any approach that relies on Dl-programs, but \cite{GiordanoDupre2016,GiordanoDupre16a} propose an ASP-based decision procedure for the DL $\mathcal{SROEL}$, relying on a Datalog encoding of the DL knowledge base.

\vspace*{-0.3cm}
\section{Conclusions} \label{concl}

\nd The introduction of rational monotonicity into the field of dl-programs allows the use of a non-monotonic formalism that at the same time satisfies important logical properties and gives back intuitive conclusions.  From the implementation point of view, our proposal allows to compile the decision procedures into dl-programs and, thus, it can be implemented on top of existing reasoners supporting dl-programs such as DLV.

\paragraph{Future work.}
We believe that two aspects are particularly urgent. Firstly, a comparison with the semantic characterisation of rational closure for \sroiq~in \cite{Bonatti19}. 

Also, we would like to address the computational complexity of our approach. So far, we know that computing the rankings can be done in polynomial number of calls (see, \eg~\cite{Casini10,CasiniEtAl2018}) to an oracle deciding \sroiq~entailment (the latter is complete for $\mathtt{2NEXP}$~\cite{Kazakov08}). It remains to be seen whether, by reasoning similarly as done in~\cite{Eiter08}, in which it has been shown that \wrt~\shoin~the existence of answer sets, cautious and brave reasoning problems are complete for $\mathtt{P}^{\mathtt{NEXP}}$,\footnote{Recall that the entailment problem for \shoin~is complete for $\mathtt{NEXP}$~\cite{Tobies01}.} the same problems are complete for $\mathtt{P}^{\mathtt{2NEXP}}$ \wrt~our \sroiq~setting, \ie~solvable in polynomial time by relying on an oracle for $\mathtt{2NEXP}$.


Eventually, from the inferential point of view rational closure has some well-known weaknesses: while there can be intuitive, desirable conclusions that cannot be derived~\cite{Lehmann95}, it remains an important basic construction that can be extended into richer entailment relations such as those proposed in~\cite{CasiniStraccia2012,CasiniStraccia13,CasiniEtAl14,GiordanoGliozzi21}. Future work will be partly dedicated to extending the present method to some of these  entailment relations.



%
%





\end{document}